\documentclass[a4paper, aps,prl,twocolumn,showpacs,superscriptaddress,groupedaddress]{revtex4}  
\usepackage{graphicx}  
\usepackage{dcolumn}   
\usepackage{bm}        
\usepackage{amsmath}
\usepackage{amssymb}   
\usepackage{enumitem}
\usepackage{graphicx,mathrsfs}
\usepackage{nicefrac}
\usepackage{multirow}
\usepackage{color}

\newcommand{\be}{\begin{equation}}  
\newcommand{\ee}{\end{equation}}  
\newcommand{\bea}{\begin{eqnarray}}  
\newcommand{\eea}{\end{eqnarray}}  

\def\gev{\, {\rm GeV}}

\DeclareRobustCommand{\fbi}{\ensuremath{\mathrm{fb}^{-1}}}
\DeclareRobustCommand{\pt}{\ensuremath{p_{\mathrm{T}}}}

\DeclareRobustCommand{\ptOne}{\ensuremath{p_{\mathrm{T}1}}}
\DeclareRobustCommand{\ptTwo}{\ensuremath{p_{\mathrm{T}2}}}

\DeclareRobustCommand{\mgg}{\ensuremath{m_{\gamma\gamma}}}
\DeclareRobustCommand{\gggg}{\ensuremath{\gamma\gamma \rightarrow \gamma\gamma}}
\DeclareRobustCommand{\baselineCoupling}{\ensuremath{\zeta_1 = 2\cdot 10^{-13}}}

\begin{document}

\title{Probing new physics in diphoton production with proton tagging at the
Large Hadron Collider} 
\author{S. Fichet}
\email{sylvain.fichet@lpsc.in2p3.fr}
\affiliation{International Institute of Physics, UFRN, Av. Odilon Gomes de Lima, 1722 - Capim Macio - 59078-400 - Natal-RN, Brazil}

\author{G. von Gersdorff}
\email{gersdorff@gmail.com}
\affiliation{ICTP SAIFR, Instituto de Fisica Teorica, Sao Paulo State University, Brazil}


\author{O. Kepka}
\email{kepkao@fzu.cz}
\affiliation{Institute of Physics of the Academy of Sciences, Prague}

\author{B. Lenzi}
\email{bruno.lenzi@cern.ch}
\affiliation{CERN, CH-1211 Geneva 23, Switzerland}

\author{C. Royon}
\email{christophe.royon@cea.fr}
\affiliation{IRFU/Service de Physique des Particules, CEA/Saclay, 91191 Gif-sur-Yvette cedex, France}

\author{M. Saimpert}
\email{matthias.saimpert@cern.ch}
\affiliation{IRFU/Service de Physique des Particules, CEA/Saclay, 91191 Gif-sur-Yvette cedex, France}

\date{\today}

\begin{abstract}
The sensitivities to anomalous quartic photon 
couplings at the Large Hadron Collider are estimated using diphoton production via photon fusion. The 
tagging of the protons proves to be a very powerful tool to suppress the background and
unprecedented sensitivities down to $6 \cdot 10^{-15}$\gev$^{-4}$ are obtained, providing a new window on extra dimensions and strongly-interacting composite states in the multi-TeV range. 
Generic contributions to quartic photon couplings from charged and neutral particles with arbitrary spin are also presented.

\end{abstract}

\maketitle


\begin{figure}
\includegraphics[scale=0.35]{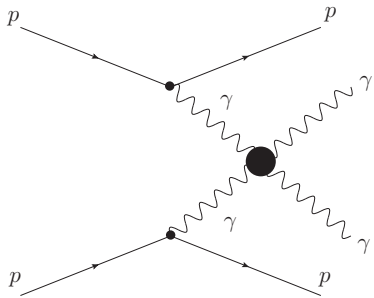}
\caption{\label{fig:diag} Diphoton production via photon fusion
sensitive to $4\gamma$ anomalous couplings.
Both protons are intact in the final state.}
\end{figure}

Several major experimental and conceptual facts, like the overwhelming
evidence for dark 
matter or the gauge-hierarchy problem, point towards the existence of 
new physics beyond the Standard Model (SM) at a scale relatively close to the 
electroweak scale.
In spite of naturalness arguments, this pradigm of a 
TeV-scale new physics is challenged by both direct searches at the
Large Hadron Collider (LHC) and by 
indirect measurements like the LEP electroweak precision tests.
In the 
scenario of new physics out of reach from direct observation at the LHC, one 
may expect that the first manifestations show up in precision measurements of 
the SM properties. Such powerful precision tests are already well advanced in 
the electroweak and flavour sectors of the SM, and distortions of the newly 
discovered Higgs sector are also being scrutinized. However, another sector of 
the SM can be tested with high precision at the LHC, the one of pure gauge 
interactions.

In this Letter, four-photon ($4\gamma$) interactions through diphoton production via photon fusion with 
intact outgoing protons are considered (Fig.~\ref{fig:diag}). Interactions between photons and $Z$, $W$ bosons in a similar case have already been studied \cite{us}.
The only existing direct limits on $4\gamma$ interactions originate from low energy laser experiments~\cite{laser}. The study of this process in LHC proton-proton collisions at center-of-mass energy of $\sqrt{s}=14$\,TeV will benefit from the new forward proton detectors considered in the ATLAS and CMS/TOTEM experiments~\cite{CMSAFP}. 
We first provide the generic new physics
contributions to the $4\gamma$ couplings and point out sizable contributions 
from strongly-coupled
and warped extra dimension scenarios. The sensitivities of the upgrades of the
ATLAS and CMS/TOTEM experiments are then given including all backgrounds.

In the assumption of a new physics mass scale $\Lambda$ heavier than experimentally accessible 
energy $E$, all new physics manifestations can be described using 
an effective Lagrangian valid for  $\Lambda\gg E$.
Among these operators, the pure photon dimension-eight operators
\be
\mathcal{L}_{4\gamma}= 
\zeta_1^\gamma F_{\mu\nu}F^{\mu\nu}F_{\rho\sigma}F^{\rho\sigma}
+\zeta_2^\gamma F_{\mu\nu}F^{\nu\rho}F_{\rho\lambda}F^{\lambda\mu}
\label{zetas}
\ee
can induce the $\gggg$ process, highly 
suppressed in the SM~\cite{Fichet:2013ola,Gupta:2011be}.
We discuss here possible new physics contributions to 
$\zeta_{1,2}^\gamma$ that can be probed and discovered at the LHC using
the forward proton detectors. 


Loops of heavy charged particles contribute to the $4\gamma$
couplings~\cite{Fichet:2013ola} as 
$\zeta_i^\gamma=\alpha^2_{\rm em} Q^4\,m^{-4}\, N\,c_{i,s}$, where
\be
c_{1,s}=
\begin{cases}
\frac{1}{288} & s=0 \\
-\frac{1}{36} & s=\frac{1}{2} \\
-\frac{5}{32} & s=1 \\
\end{cases}
\,,\quad
c_{2,s}=
\begin{cases}
\frac{1}{360} & s=0 \\
\frac{7}{90} & s=\frac{1}{2} \\
\frac{27}{40} & s=1 \\
\end{cases}
\ee
where $s$ denotes the spin of the heavy particle of mass $m$ running in the loop and $Q$ 
its electric charge. 
The factor $N$  counts all additional multiplicities such as color or flavor.
These couplings scale as $\sim Q^4$ and are enhanced 
in presence of  particles with large charges. For example, certain light 
composite fermions, characteristic of composite Higgs models, have typically 
electric charges of several units~\cite{Fichet:2013ola}. For a 500\gev vector (fermion) resonance with  
$Q=3\ (4)$,
large couplings $\zeta_i^{\gamma}$ of the order of
$10^{-13}-10^{-14}$\gev$^{-4}$ can be reached.

Beyond perturbative contributions to $\zeta_i^\gamma$ from  
charged particles,  non-renormalizable interactions of neutral particles are 
also present in common extensions of the SM.  Such theories can contain 
scalar, pseudo-scalar and spin-2 resonances, respectively denoted $\varphi$, 
$\tilde \varphi$, $h^{\mu\nu}$, that couple  to the photon 
as
\be\begin{split}
\mathcal L_{\gamma\gamma}=&f_{0^+}^{-1}\,\varphi\, 
(F_{\mu\nu})^2+f_{0^-}^{-1}\, \tilde\varphi \, F_{\mu\nu}F_{\rho\lambda}\,
\epsilon^{\mu\nu\rho\lambda} \\&+f_{2}^{-1}\, h^{\mu\nu}\, (-F_{\mu\rho} 
F_{\nu}^{\,\,\rho}+\eta_{\mu\nu} (F_{\rho\lambda})^2/4)\,,
\end{split}
\ee
and generate  the $4\gamma$ couplings by tree-level exchange as 
$\zeta_i^\gamma=(f_{s}\, m)^{-2}\,d_{i, s}$, where
\be
d_{1,s}=
\begin{cases}
\frac{1}{2} & s=0^+ \\
 -4 & s=0^- \\
-\frac{1}{8} & s=2\\
\end{cases}
\,,\quad
d_{2,s}=
\begin{cases}
0 & s=0^+ \\
8  & s=0^- \\
\frac{1}{2} & s=2 \\
\end{cases}
\,.
\ee

Strongly-coupled conformal extensions of the SM contain a scalar particle $(s=0^+)$, 
the dilaton. 
In the case of small explicit conformal breaking, the dilaton is light and couples only weakly to the photon, 
$f_{\varphi}^{-1}\ll m_\varphi^{-1}$.
However, a more natural situation occurs when explicit conformal breaking is large \cite{Chacko:2012sy,Bellazzini:2012vz,Chacko:2013dra}, in which case the dilaton has a mass comparable to the other resonances of the theory and can be much more stongly coupled
$f_\varphi^{-1}\sim \pi/m_{\varphi}$,
as long as photons are mostly composite.
In this case, even a 2 TeV dilaton can produce a sizable effective photon interaction, $\zeta_1^\gamma\sim 10^{-13}$\gev$^{-4}$.

These features are reproduced at large number of colours by the 
gauge-gravity correspondence in a warped extra dimension. The 
dilaton is identified as the radion,  and a mainly composite photon 
corresponds to a large infrared (IR) brane kinetic term.
Warped-extra dimensions also feature Kaluza-Klein (KK) gravitons \cite{Randall:1999ee}.
These are interpreted as spin 2 resonances in the gauge theory. A mostly 
elementary photon does not yield a sizeable coupling. However, a mostly
composite one couples more strongly to the KK fields, in that case the 
whole set of KK modes induces \cite{Fichet:2013ola}
 \be
\zeta_i^\gamma= \frac{\kappa^2}{8 \tilde k^4} \,d_{i, 2}
\,,\ee 
where $\tilde k$ is the IR scale that determines the first KK graviton mass 
as $m_{2}=3.83\, \tilde k$, and $\kappa$ is a parameter that can be taken 
$\mathcal O(1)$. 
For $\kappa\sim 1$, and $m_{2}\lesssim 6$ TeV, the photon vertex can easily 
exceed $\zeta_2^\gamma\sim 10^{-14}$\gev$^{-4}$.
 
Since we deal with non-renormalizable couplings
perturbative unitarity (and effective field theory) breaks down 
at some scale $\Lambda'$. This can partially be avoided by using 
full amplitudes, but even then some couplings (such as the dilaton coupling 
$f_\varphi^{-1}$) grow with energy. Whenever the scale $\Lambda'$ falls below the 
detector acceptance a form factor $1/(1+(m_{\gamma\gamma}^{2}/\Lambda')^2)$ is applied to mimic the 
effects that restore unitarity~\cite{Gupta:2011be}.
In many cases such a form factor is not necessary (for instance, when the new particles have a large enough mass).
However, for completeness sensitivities with $\Lambda' = 1$ TeV are quoted.

The \gggg\ process (Fig.~\ref{fig:diag}) can be probed via the detection of 
two intact protons in the forward proton detectors proposed by the 
ATLAS and CMS~\cite{CMSAFP} collaborations, and two 
energetic photons in the corresponding electromagnetic 
calorimeters~\cite{ATLASdetectorpaper,CMSdetectorpaper}. The 
forward detectors are expected to be located symmetrically at about 210~m from the main 
interaction point and cover the range $0.015 < \xi < 0.15$, 
where $\xi$ is the fractional proton momentum loss. The time-of-flight of the scattered proton can be measured with a precision of $\sim 10$~ps that allows 
to determine the production point of the protons within 2.1~mm inside ATLAS/CMS and to check 
if they originate from the same scattering vertex as the two photons. It is worth noticing
that the SM cross section of diphoton production with intact protons
is dominated by the
QED process at high diphoton mass --- and not by gluon exchanges --- and is thus very well known
. If the protons 
are not intact, the two-photon quasi-elastic diphoton production with large theoretical 
uncertainties needs to be considered~\cite{szczurek}, leading
to a large uncertainty in the background determination. In the present case, any deviation from the
standard model prediction will be a sign of beyond SM effects.

The electromagnetic calorimeters cover the pseudorapidity range 
$|\eta| \lesssim 2.5$ and provide excellent resolution in terms of energy ($<1\%$ at transverse momenta $\pt > 100$~GeV) and 
position (0.001 in $\eta$ and 1 mrad in the azimuthal 
angle $\phi$) for photons with \pt\ ranging from few\gev to 
few TeV~\cite{ATLASCSC}. A fraction of the photons ($\sim 15-30$\%) converts to 
electron-positron pairs in the region instrumented with silicon 
tracking detectors. The reconstruction of at least one conversion allows to locate the 
photon production point with sub-millimiter accuracy and, when combined 
with the information from the proton detectors, constrains the full event 
kinematics. This is extremely powerful to reject 
backgrounds with real or fake photons from a hard scattering process and 
protons coming from additional interactions occurring in the same or 
neighbouring bunch crossings (pile-up). The average number of multiple proton-proton collisions per 
bunch crossing is denoted as $\mu$ in the following. In the case of the ATLAS detector, the production point of the photons can be determined within $\sim 15$~mm exploiting the longitudinal segmentation of
the ATLAS calorimeter~\cite{ATLASHiggsobservation}. Consequently, an alternative scenario with no
converted photons is also considered.

According to Ref.~\cite{ATLASECFA}, even in the presence of more than 
100 pile-up interactions, the photon identification efficiency is 
expected to be around 75\% for $\pt > 100$~GeV, with jet rejection 
factors exceeding 4000. In addition, about 1\% of the electrons are 
mis-identified as photons. These numbers are used in the phenomenological
study presented below.

\begin{figure}
\includegraphics[scale=0.35]{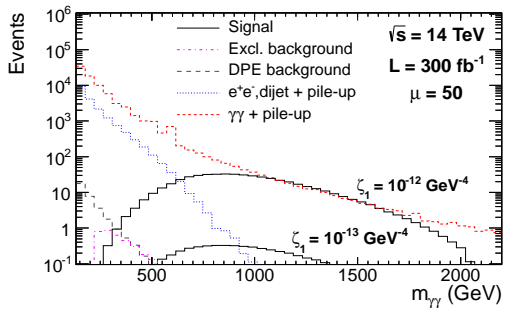}
\caption{ Diphoton invariant
mass distribution for the signal ($\zeta_{1} = 10^{-12},~10^{-13}$\gev$^{-4}$, see Eq.~\ref{zetas}) and for the backgrounds (dominated by $\gamma\gamma$ with protons from pile-up), requesting
two protons in the forward detectors and two photons of $\pt > 50$~GeV with at least one converted photon in the central detector, for a luminosity of 300$~\fbi$ and an average pile-up of $\mu = 50$. Excl. stands for exclusive backgrounds and DPE for double pomeron exchange backgrounds (see text).}
\label{fig:mass}
\end{figure}

\begin{table*}
\small
\caption{Number of signal (for \baselineCoupling\gev$^{-4}$) and background events after 
various selections for an integrated
luminosity of 300~\fbi\ and $\mu=50$ at $\sqrt{s}=14$ TeV. At least one converted photon is required. Excl. stands for exclusive backgrounds and DPE for double pomeron exchange backgrounds (see text).}
\begin{tabular}{|c||c||c|c|c|c|}
\hline
Cut / Process & Signal & Excl. & DPE & e$^{+}$e$^{-}$, dijet + pile-up  & $\gamma\gamma$ + pile-up \\
\hline
\hline
$0.015<\xi<0.15$, $p_{\mathrm{T}1,2}>50$~GeV                  & 20.8  & 3.7 & 48.2 & $2.8\cdot10^{4}$ & $1.0\cdot10^{5}$ \\
$p_{\mathrm{T}1}>200$GeV, $p_{\mathrm{T}2}>100$~GeV                    & 17.6  & 0.2 & 0.2  & 1.6          & 2968         \\
$m_{\gamma\gamma}>600$~GeV                          & 16.6  & 0.1 &  0    & 0.2          & 1023         \\
$p_{\mathrm{T2}}/p_{\mathrm{T1}}>0.95$, $|\Delta\phi|>\pi-0.01$      & 16.2  & 0.1 & 0   & 0          & 80.2         \\
$\sqrt{\xi_{1}\xi_{2}s} = m_{\gamma\gamma} \pm 3\%$                                      & 15.7  & 0.1 & 0   & 0          & 2.8          \\
$|y_{\gamma\gamma}-y_{pp}|<0.03$                   & 15.1  & 0.1 & 0   & 0            & 0            \\
\hline

\end{tabular}
\label{tab:event}
\end{table*}

The anomalous \gggg\ process has been 
implemented in the Forward Physics Monte Carlo (FPMC) generator~\cite{FPMC}
that aims at providing a variety of diffractive and photon-induced processes in 
a common framework, including a survival probability of 0.9
. This factor is necessary to take into account the possibility of additional soft interactions occurring between the two intact protons. The FPMC generator was used to simulate the signal and background 
processes giving rise to two intact protons accompanied by two photons, 
electrons or jets that can mimic the photon signal. Those include exclusive 
SM production of \gggg\ via lepton and quark boxes and 
$\gamma\gamma\rightarrow e^{+}e^{-}$. The central exclusive 
production of $\gamma\gamma$ via two-gluon exchange, not present in FPMC,
was simulated using ExHuME~\cite{Monk:2005ji}. This series of backgrounds is called
``Exclusive" in Table~\ref{tab:event} and Figs.~2, 3.
FPMC was also used to produce $\gamma\gamma$, 
Higgs to $\gamma\gamma$ and dijet productions via double pomeron exchange (called
DPE background in Table~\ref{tab:event} and Fig.~2).
Such backgrounds tend to be softer than the signal and can be suppressed with 
requirements on the transverse momenta of the photons ($\ptOne > 200$~GeV 
for the leading and $\ptTwo > 100$~GeV for the subleading photons, respectively) and the diphoton invariant 
mass ($\mgg > 600$~GeV), as shown in Fig.~\ref{fig:mass}. In addition, the final-state photons of the signal are typically back-to-back and have 
about the same transverse momenta. Requiring a large azimuthal 
angle $|\Delta \phi| > \pi -0.01$ between the two photons and a 
ratio $\ptTwo / \ptOne > 0.95$ greatly reduces the contribution of non-exclusive processes.

\begin{figure*}
\includegraphics[scale=0.35]{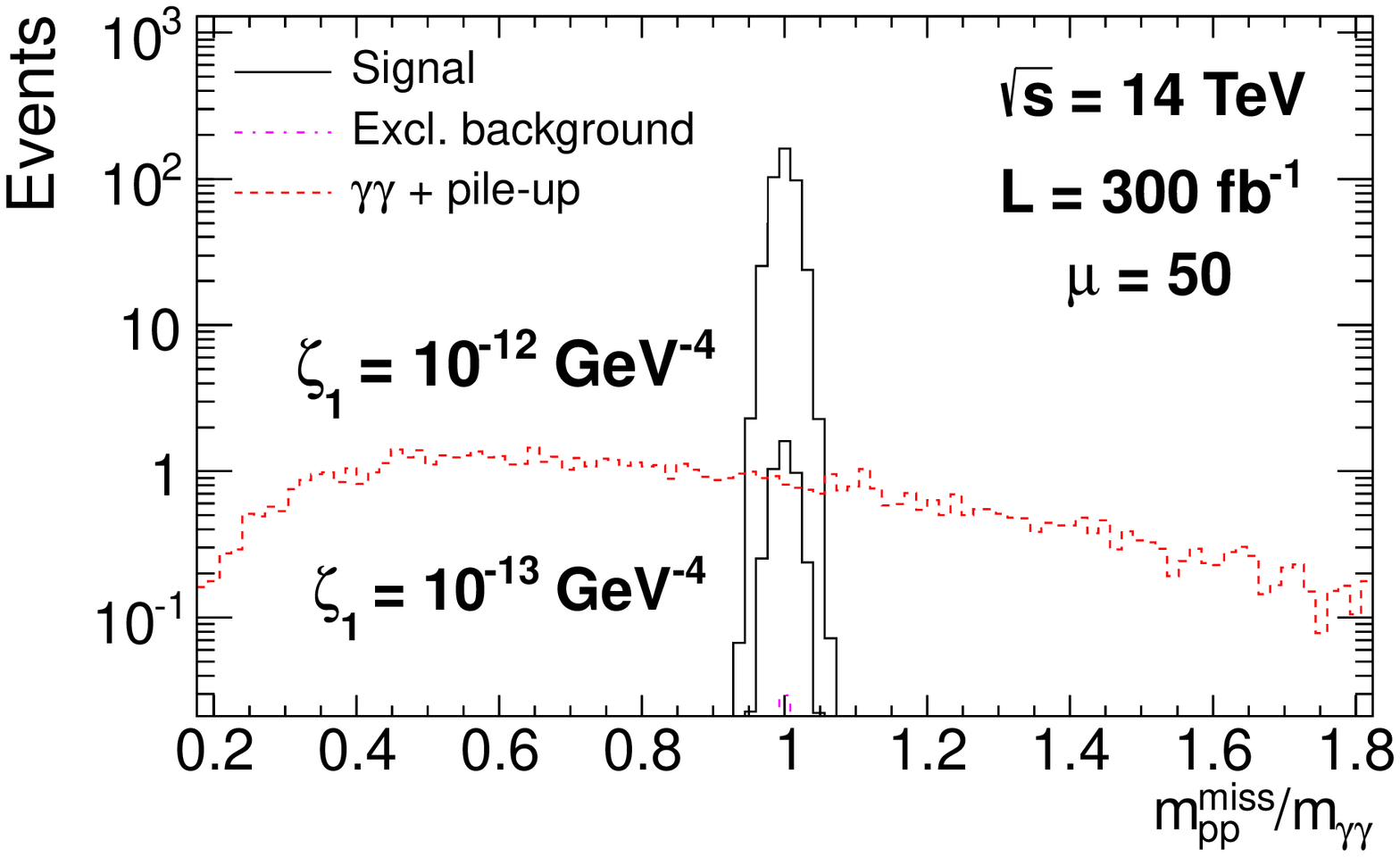}
\includegraphics[scale=0.35]{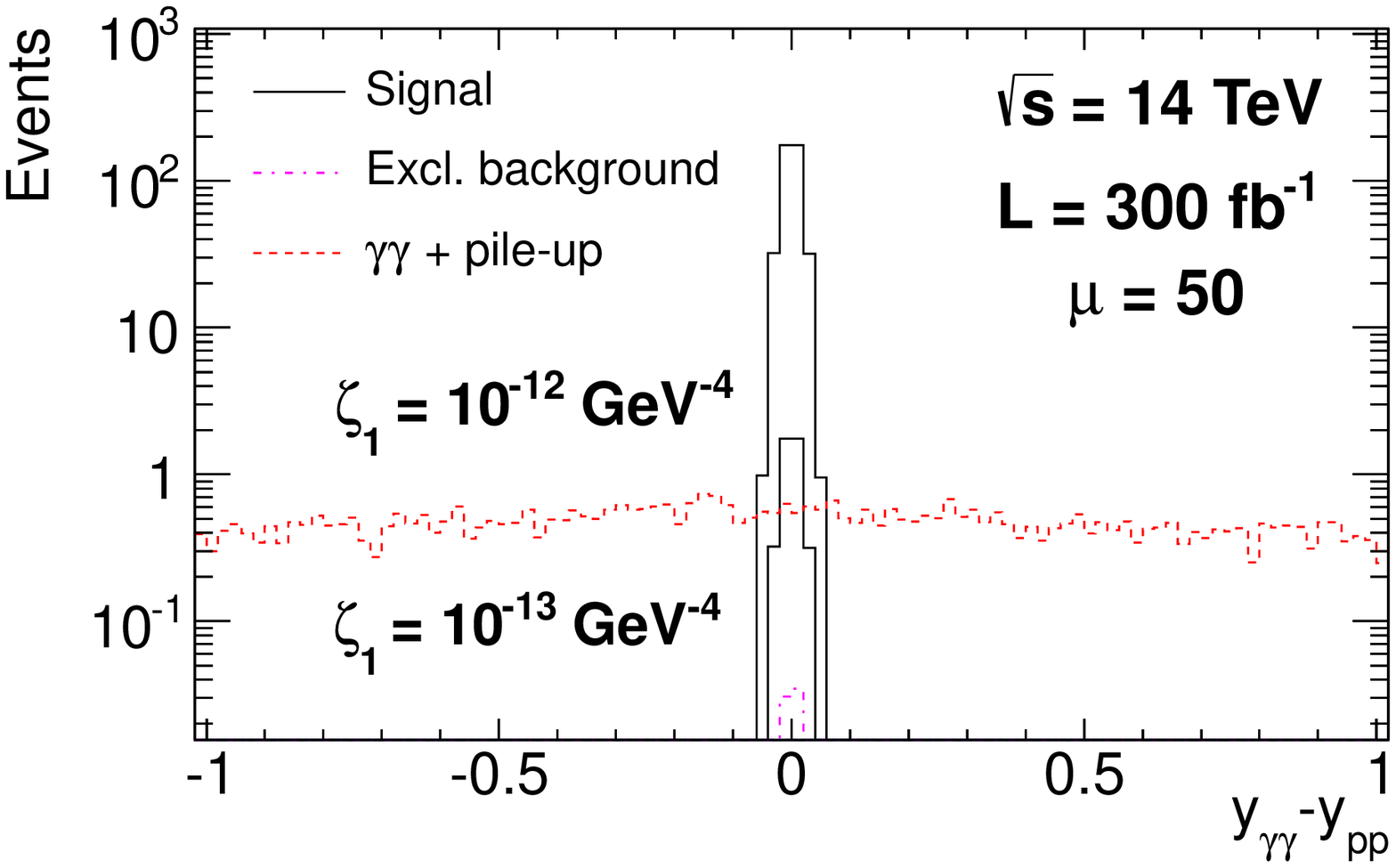}
\caption{\label{fig:massratio} Diphoton to missing proton mass ratio (left) and rapidity difference (right)
distributions for signal considering two 
different coupling values ($10^{-12}$ and $10^{-13}$\gev$^{-4}$, see Eq.~\ref{zetas}) and for 
backgrounds after requirements on photon \pt, diphoton invariant mass, \pt\ ratio between the two photons and on the angle between the two photons. At least one converted photon is required. The integrated luminosity 
is 300~\fbi\ and the average pile-up is $\mu=50$.}
\end{figure*}

Additional background processes include the quark and gluon-initiated 
production of two photons, two jets and Drell-Yan processes leading to two electrons. The two intact 
protons arise from pile-up interactions (these backgrounds are called $\gamma\gamma$ + pile-up and e$^{+}$e$^{-}$, dijet + pile-up in Table~\ref{tab:event}). The hard scattering processes are 
simulated with the HERWIG 6.5~\cite{Corcella:2002jc} generator while the pile-up 
interactions are simulated by PYTHIA8~\cite{PYTHIA8}. The probability to detect at least 
one proton in each of the two forward detectors is estimated to be 32\%, 
66\% and 93\% for 50, 100 and 200 additional interactions, respectively. 
The pile-up background is further suppressed by requiring the proton missing invariant mass to match the diphoton invariant mass within the expected 
resolution ($m_{pp}^{\mathrm{miss}}=\sqrt{\xi_1 \xi_2 s}=m_{\gamma\gamma} \pm 3\%$), and the diphoton
system rapidity and the rapidity of the two protons defined as $y_{pp}=0.5$~ln($\frac{\xi_1}{\xi_2})$ to be the same within the resolution
($|y_{\gamma \gamma} - y_{pp}| < 0.03$),
as shown in 
Fig.~\ref{fig:massratio}. 

The number of expected signal and background events passing respective selections is shown in 
Table~\ref{tab:event} for an integrated luminosity of 300~\fbi\ ($\simeq$ 3 years of data-taking at the LHC) and 50 pile-up interactions for a center-of-mass energy of 14\,TeV.
It is required that at least one photon converts in the tracker. Gaussian smearings of 1\% for the total energy, 0.001 for the pseudorapidity and 1~mrad 
for the azimuthal angle are applied to each photon.
Exploiting the full event kinematics with the forward proton detectors 
allows to completely suppress the background with a signal selection 
efficiency after the acceptance cuts exceeding 70\%. Tagging the protons
is absolutely needed to suppress the $\gamma \gamma$ + pile-up events.
Further background reduction is even possible by requiring the photons 
and the protons to originate from the same vertex that provides an additional rejection 
factor of 40 for 50 pile-up interactions, showing the large margin on the background suppression.
A similar study 
at a higher pile-up of 200 was performed 
and led to a negligible background (0.3 
expected background events for 300~\fbi), showing the robustness of this 
analysis. 
Moreover, if one relaxes the request of at least one photon to be converted, 
the signal is increased by a factor 3 to 4.
The sensitivities
on photon quartic anomalous couplings are given in Table~\ref{sensitivities}
for different scenarios corresponding to the medium luminosity at the LHC
(300 fb$^{-1}$) and the high luminosity (6000 fb$^{-1}$ when combining the two experiments ATLAS and CMS/TOTEM). 
The sensitivity extends up to $6\cdot10^{-15}$ allowing us to probe further the models of new
physics described above.


\begin{table}
\caption{$5\sigma$ discovery and 95\% CL exclusion limits on $\zeta_1$ and $\zeta_2$ 
couplings in\gev$^{-4}$ (see Eq.~\ref{zetas}) with
and without form factor (f.f.), requesting at least one converted photon ($\ge$~1 conv. $\gamma$) or
not (all $\gamma$). All sensitivities are given for 300 fb$^{-1}$
and $\mu=50$ pile-up events (medium luminosity LHC) except for the numbers of the last column which are given for 6000
fb$^{-1}$ and $\mu=200$ pile-up events (high luminosity LHC).}
\label{sensitivities}

\begin{tabular}{|c||c|c||c||c|}
\hline
Luminosity & 300~\fbi & 300~\fbi & 300~\fbi & 6000~\fbi \\
\hline
pile-up ($\mu$) & 50 & 50 & 50 & 200 \\
\hline
\hline
coupling & $\ge$~1 conv. $\gamma$ & $\ge$~1 conv. $\gamma$ & all $\gamma$  & all $\gamma$\\
(GeV$^{-4}$) & 5 $\sigma$ & 95\% CL & 95\% CL & 95\% CL \\

\hline
$\zeta_1$~f.f. &  $1\cdot10^{-13}$ & $7\cdot10^{-14}$ & $4\cdot 10^{-14}$ & $2\cdot10^{-14}$ \\
$\zeta_1$~no f.f.& $3\cdot10^{-14}$ & $2\cdot10^{-14}$ & $1\cdot10^{-14}$ & $6\cdot10^{-15}$\\
\hline
$\zeta_2$~f.f. & $3\cdot10^{-13}$ & $1.5\cdot10^{-13}$ & $8\cdot10^{-14}$ & $4\cdot10^{-14}$ \\
$\zeta_2$~no f.f.& $7\cdot10^{-14}$ & $2\cdot10^{-14}$ & $2\cdot10^{-14}$ & $1\cdot10^{-14}$ \\
\hline

\end{tabular}
\end{table}

In this Letter, the  sensitivities to quartic photon  couplings at  the LHC,  
obtained by  measuring the photons in the central CMS and ATLAS detectors and the
intact protons in dedicated forward proton detectors, are estimated. 
For the first time, sensitivities on anomalous quartic couplings are large enough to probe 
models of new physics. 
The imprint of warped KK gravitons and of a strongly-coupled dilaton can be discovered in the multi-TeV range. Also, a generic 500\gev fermion (vector) resonance can be probed for electric charge $Q\gtrsim 4$ $(3)$ via loop effects.
%
The analysis greatly benefits from the kinematical
constraints from the photon and proton measurements, which allows us to obtain 
negligible backgrounds. 

We thank useful discussions with Christophe Grojean.


\begin{thebibliography}{99}


\bibitem{us} 
  E. Chapon, O. Kepka, C. Royon, Phys. Rev. {\bf D81} (2010) 074003;
  O.~Kepka and C.~Royon,
  Phys.\ Rev.\  D {\bf 78} (2008) 073005;
  J. de. Favereau et al., preprint arXiv:0908.2020.

\bibitem{laser}
  M. Bregant et {\it{al}},
  Phys.\ Rev.\ D {\bf{78}} (2008) 032006.

\bibitem{CMSAFP}
ATLAS Coll., CERN-LHCC-2011-012; TOTEM Coll., CERN-LHCC-2013-009.  


\bibitem{Fichet:2013ola}
  S.~Fichet and G.~von Gersdorff,
  preprint arXiv:1311.6815.

\bibitem{Gupta:2011be}
  R.~S.~Gupta,
  Phys.\ Rev.\ D {\bf 85} (2012) 014006.


\bibitem{Chacko:2012sy} 
  Z.~Chacko and R.~K.~Mishra,
  Phys.\ Rev.\ D {\bf 87} (2013) 115006.

\bibitem{Bellazzini:2012vz} 
  B.~Bellazzini, et al., 
  Eur.\ Phys.\ J.\ C {\bf 73} (2013) 2333.


\bibitem{Chacko:2013dra} 
  Z.~Chacko, R.~K.~Mishra and D.~Stolarski,
  JHEP {\bf 1309} (2013) 121.

\bibitem{Randall:1999ee} 
  L.~Randall and R.~Sundrum,
  Phys.\ Rev.\ Lett.\  {\bf 83} (1999) 3370.



\bibitem{ATLASdetectorpaper} ATLAS Coll., JINST, Vol. 3 (2008) S08003.

\bibitem{CMSdetectorpaper} CMS Coll., JINST, Vol. 3 (2008) S08004.


\bibitem{szczurek} M. Luszczak, A. Szczurek, preprint arXiv:1309.7201.


\bibitem{ATLASCSC} ATLAS Coll., http://cdsweb.cern.ch/record/1125884, CERN-OPEN-2008-020.

\bibitem{ATLASHiggsobservation} ATLAS Coll., Phys. Lett. B {\bf 716} (2012) 1.

\bibitem{ATLASECFA} ATLAS Coll., ATL-PHYS-PUB-2013-009.

\bibitem{FPMC}  
  M. Boonekamp et al.,
preprint  arXiv:1102.2531.




\bibitem{Monk:2005ji}
  J.~Monk and A.~Pilkington,
  Comput.\ Phys.\ Commun.\  {\bf 175} (2006) 232; V.A. Khoze, A.D. Martin, M.G. Ryskin,
  Eur.Phys.J. C {\bf 55} (2008) 363. 

\bibitem{Corcella:2002jc} G. Corcella et al., arXiv:hep-ph/0210213.


\bibitem{PYTHIA8}
  T.~Sjostrand, S.~Mrenna and P.~Z.~Skands,
  Comput.\ Phys.\ Commun.\  {\bf 178} (2008) 852.
  








\end{thebibliography}

\end{document}